\pdfoutput=1

\documentclass[12pt,a4paper]{article}

\usepackage{ifthen} 
\newboolean{pdflatex}
\setboolean{pdflatex}{true} 

\newboolean{articletitles}
\setboolean{articletitles}{true} 

\newboolean{uprightparticles}
\setboolean{uprightparticles}{false} 

\newboolean{inbibliography}
\setboolean{inbibliography}{false} 

\def\paperauthors{LHCb collaboration} 
\def\papertitle{Search for the rare decay \Bmumumu}
\def\paperasciititle{Search for the rare decay Bplus to muplus muminus neutrino} 
\def\paperkeywords{{High Energy Physics}, {LHCb}, {leptonic decay}, {Vub}}
\def\papercopyright{\the\year\ CERN for the benefit of the LHCb collaboration} 
\def\paperlicence{CC-BY-4.0 licence}
\def\paperlicenceurl{https://creativecommons.org/licenses/by/4.0/}

\usepackage[top=1in, bottom=1.25in, left=1in, right=1in]{geometry}

\usepackage{microtype}
\usepackage{lineno}  
\usepackage{xspace} 
\usepackage{caption} 

\usepackage{graphicx}  
\usepackage{color}
\usepackage{colortbl}
\graphicspath{{./figs/}} 

\usepackage{amsmath} 
\usepackage{amssymb}
\usepackage{amsfonts}
\usepackage{upgreek} 

\newcommand*\patchAmsMathEnvironmentForLineno[1]{%
\expandafter\let\csname old#1\expandafter\endcsname\csname #1\endcsname
\expandafter\let\csname oldend#1\expandafter\endcsname\csname
end#1\endcsname
 \renewenvironment{#1}%
   {\linenomath\csname old#1\endcsname}%
   {\csname oldend#1\endcsname\endlinenomath}%
}
\newcommand*\patchBothAmsMathEnvironmentsForLineno[1]{%
  \patchAmsMathEnvironmentForLineno{#1}%
  \patchAmsMathEnvironmentForLineno{#1*}%
}
\AtBeginDocument{%
\patchBothAmsMathEnvironmentsForLineno{equation}%
\patchBothAmsMathEnvironmentsForLineno{align}%
\patchBothAmsMathEnvironmentsForLineno{flalign}%
\patchBothAmsMathEnvironmentsForLineno{alignat}%
\patchBothAmsMathEnvironmentsForLineno{gather}%
\patchBothAmsMathEnvironmentsForLineno{multline}%
\patchBothAmsMathEnvironmentsForLineno{eqnarray}%
}

\usepackage{hyperxmp}

\usepackage[pdftex,
            pdfauthor={\paperauthors},
            pdftitle={\paperasciititle},
            pdfkeywords={\paperkeywords},
            pdfcopyright={Copyright (C) \papercopyright},
            pdflicenseurl={\paperlicenceurl}]{hyperref}

\usepackage[all]{hypcap} 

\usepackage{xspace} 
\usepackage{upgreek}

\def\lhcb {\mbox{LHCb}\xspace}

\def\MagUp {\mbox{\em Mag\kern -0.05em Up}\xspace}

\ifthenelse{\boolean{uprightparticles}}%
{

 \def\Pmu         {\ensuremath{\upmu}\xspace}                 
 \def\Pnu         {\ensuremath{\upnu}\xspace}                 
                  
 \def\Ppi         {\ensuremath{\uppi}\xspace}

 \def\Ptau        {\ensuremath{\uptau}\xspace}

 \def\Ppsi        {\ensuremath{\uppsi}\xspace}

 \def\PDelta      {\ensuremath{\Delta}\xspace}                 
 \def\PXi      {\ensuremath{\Xi}\xspace}                 
 \def\PLambda      {\ensuremath{\Lambda}\xspace}                 
 \def\PSigma      {\ensuremath{\Sigma}\xspace}                 
 \def\POmega      {\ensuremath{\Omega}\xspace}                 
 \def\PUpsilon      {\ensuremath{\Upsilon}\xspace}                 
 
 \def\PB      {\ensuremath{\mathrm{B}}\xspace}                 
                  
 \def\PD      {\ensuremath{\mathrm{D}}\xspace}

 \def\PJ      {\ensuremath{\mathrm{J}}\xspace}                 
 \def\PK      {\ensuremath{\mathrm{K}}\xspace}

 \def\Pb      {\ensuremath{\mathrm{b}}\xspace}                 
 \def\Pc      {\ensuremath{\mathrm{c}}\xspace}

 \def\Pi      {\ensuremath{\mathrm{i}}\xspace}

}
{

 \def\Pmu         {\ensuremath{\mu}\xspace}                 
 \def\Pnu         {\ensuremath{\nu}\xspace}                 
                  
 \def\Ppi         {\ensuremath{\pi}\xspace}

 \def\Ptau        {\ensuremath{\tau}\xspace}

 \def\Ppsi        {\ensuremath{\psi}\xspace}                 
                  
 \mathchardef\PDelta="7101
 \mathchardef\PXi="7104
 \mathchardef\PLambda="7103
 \mathchardef\PSigma="7106
 \mathchardef\POmega="710A
 \mathchardef\PUpsilon="7107
                  
 \def\PB      {\ensuremath{B}\xspace}                 
                  
 \def\PD      {\ensuremath{D}\xspace}

 \def\PJ      {\ensuremath{J}\xspace}                 
 \def\PK      {\ensuremath{K}\xspace}

 \def\Pb      {\ensuremath{b}\xspace}                 
 \def\Pc      {\ensuremath{c}\xspace}

 \def\Pi      {\ensuremath{i}\xspace}

}

\makeatletter
\ifcase \@ptsize \relax
  \newcommand{\miniscule}{\@setfontsize\miniscule{4}{5}}
\or
  \newcommand{\miniscule}{\@setfontsize\miniscule{5}{6}}
\or
  \newcommand{\miniscule}{\@setfontsize\miniscule{5}{6}}
\fi
\makeatother

\DeclareRobustCommand{\optbar}[1]{\shortstack{{\miniscule (\rule[.5ex]{1.25em}{.18mm})}
  \\ [-.7ex] $#1$}}

\def\mup        {{\ensuremath{\Pmu^+}}\xspace}
\def\mun        {{\ensuremath{\Pmu^-}}\xspace} 
\def\mumu       {{\ensuremath{\Pmu^+\Pmu^-}}\xspace}

\def\taup       {{\ensuremath{\Ptau^+}}\xspace}

\def\neu        {{\ensuremath{\Pnu}}\xspace}

\def\neum       {{\ensuremath{\neu_\mu}}\xspace}

\def\neut       {{\ensuremath{\neu_\tau}}\xspace}

\def\cquark    {{\ensuremath{\Pc}}\xspace}

\def\bquark    {{\ensuremath{\Pb}}\xspace}

\def\pion   {{\ensuremath{\Ppi}}\xspace}

\def\pip    {{\ensuremath{\pion^+}}\xspace}
\def\pim    {{\ensuremath{\pion^-}}\xspace}

\def\kaon    {{\ensuremath{\PK}}\xspace}
  \def\Kbar    {{\kern 0.2em\overline{\kern -0.2em \PK}{}}\xspace}

\def\KorKbar    {\kern 0.18em\optbar{\kern -0.18em K}{}\xspace}

\def\Kp      {{\ensuremath{\kaon^+}}\xspace}

\def\Kstarz  {{\ensuremath{\kaon^{*0}}}\xspace}

  \def\Dbar    {{\kern 0.2em\overline{\kern -0.2em \PD}{}}\xspace}

\def\DorDbar    {\kern 0.18em\optbar{\kern -0.18em D}{}\xspace}

\def\Dzb     {{\ensuremath{\Dbar{}^0}}\xspace}

\def\B       {{\ensuremath{\PB}}\xspace}
\def\Bbar    {{\ensuremath{\kern 0.18em\overline{\kern -0.18em \PB}{}}}\xspace}

\def\BorBbar    {\kern 0.18em\optbar{\kern -0.18em B}{}\xspace}
\def\Bz      {{\ensuremath{\B^0}}\xspace}

\def\Bu      {{\ensuremath{\B^+}}\xspace}

\def\Bp      {{\ensuremath{\Bu}}\xspace}

\def\jpsi     {{\ensuremath{{\PJ\mskip -3mu/\mskip -2mu\Ppsi\mskip 2mu}}}\xspace}
\def\psitwos  {{\ensuremath{\Ppsi{(2S)}}}\xspace}

  \def\Y#1S{\ensuremath{\PUpsilon{(#1S)}}\xspace}

\def\Lbar        {{\ensuremath{\kern 0.1em\overline{\kern -0.1em\PLambda}}}\xspace}
\def\LorLbar    {\kern 0.18em\optbar{\kern -0.18em \PLambda}{}\xspace}

\def\BF         {{\ensuremath{\mathcal{B}}}\xspace}

\newcommand{\decay}[2]{\ensuremath{#1\!\to #2}\xspace}         

\def\to                 {\ensuremath{\rightarrow}\xspace}

\def\AT#1     {\ensuremath{A_{\mathrm{T}}^{#1}}\xspace}           

\def\C#1      {\ensuremath{\mathcal{C}_{#1}}\xspace}                       
\def\Cp#1     {\ensuremath{\mathcal{C}_{#1}^{'}}\xspace}                    
\def\Ceff#1   {\ensuremath{\mathcal{C}_{#1}^{\mathrm{(eff)}}}\xspace}        
\def\Cpeff#1  {\ensuremath{\mathcal{C}_{#1}^{'\mathrm{(eff)}}}\xspace}       
\def\Ope#1    {\ensuremath{\mathcal{O}_{#1}}\xspace}                       
\def\Opep#1   {\ensuremath{\mathcal{O}_{#1}^{'}}\xspace}                    

\newcommand{\tev}{\ifthenelse{\boolean{inbibliography}}{\ensuremath{~T\kern -0.05em eV}}{\ensuremath{\mathrm{\,Te\kern -0.1em V}}}\xspace}
\newcommand{\gev}{\ensuremath{\mathrm{\,Ge\kern -0.1em V}}\xspace}
\newcommand{\mev}{\ensuremath{\mathrm{\,Me\kern -0.1em V}}\xspace}
\newcommand{\kev}{\ensuremath{\mathrm{\,ke\kern -0.1em V}}\xspace}
\newcommand{\ev}{\ensuremath{\mathrm{\,e\kern -0.1em V}}\xspace}
\newcommand{\gevc}{\ensuremath{{\mathrm{\,Ge\kern -0.1em V\!/}c}}\xspace}
\newcommand{\mevc}{\ensuremath{{\mathrm{\,Me\kern -0.1em V\!/}c}}\xspace}
\newcommand{\gevcc}{\ensuremath{{\mathrm{\,Ge\kern -0.1em V\!/}c^2}}\xspace}
\newcommand{\gevgevcccc}{\ensuremath{{\mathrm{\,Ge\kern -0.1em V^2\!/}c^4}}\xspace}
\newcommand{\mevcc}{\ensuremath{{\mathrm{\,Me\kern -0.1em V\!/}c^2}}\xspace}

\def\mum  {\ensuremath{{\,\upmu\mathrm{m}}}\xspace}

\def\invfb   {\ensuremath{\mbox{\,fb}^{-1}}\xspace}

\newcommand{\chisq}{\ensuremath{\chi^2}\xspace}

\newcommand{\chisqip}{\ensuremath{\chi^2_{\text{IP}}}\xspace}

\def\gsim{{~\raise.15em\hbox{$>$}\kern-.85em
          \lower.35em\hbox{$\sim$}~}\xspace}
\def\lsim{{~\raise.15em\hbox{$<$}\kern-.85em
          \lower.35em\hbox{$\sim$}~}\xspace}

\def\ptot       {\mbox{$p$}\xspace}
\def\pt         {\mbox{$p_{\mathrm{ T}}$}\xspace}

\def\evtgen     {\mbox{\textsc{EvtGen}}\xspace}

\def\geant      {\mbox{\textsc{Geant4}}\xspace}

\def\photos     {\mbox{\textsc{Photos}}\xspace}

\def\pythia     {\mbox{\textsc{Pythia}}\xspace}

\def\tell1  {TELL1\xspace}
\def\ukl1   {UKL1\xspace}

\usepackage{todonotes}

\usepackage{cite} 
\usepackage{mciteplus}

\def\Bmumumu {\decay{\Bp}{\mup\mun\mup\neum}}

\def\bjpsimumuk  {\decay{\Bp}{\ensuremath{{\jpsi}\decay({\mup\mun})\Kp}}}

\begin{document}

\renewcommand{\thefootnote}{\fnsymbol{footnote}}
\setcounter{footnote}{1}


\begin{titlepage}
\pagenumbering{roman}

\vspace*{-1.5cm}
\centerline{\large EUROPEAN ORGANIZATION FOR NUCLEAR RESEARCH (CERN)}
\vspace*{1.5cm}
\noindent
\begin{tabular*}{\linewidth}{lc@{\extracolsep{\fill}}r@{\extracolsep{0pt}}}
\ifthenelse{\boolean{pdflatex}}
{\vspace*{-1.5cm}\mbox{\!\!\!\includegraphics[width=.14\textwidth]{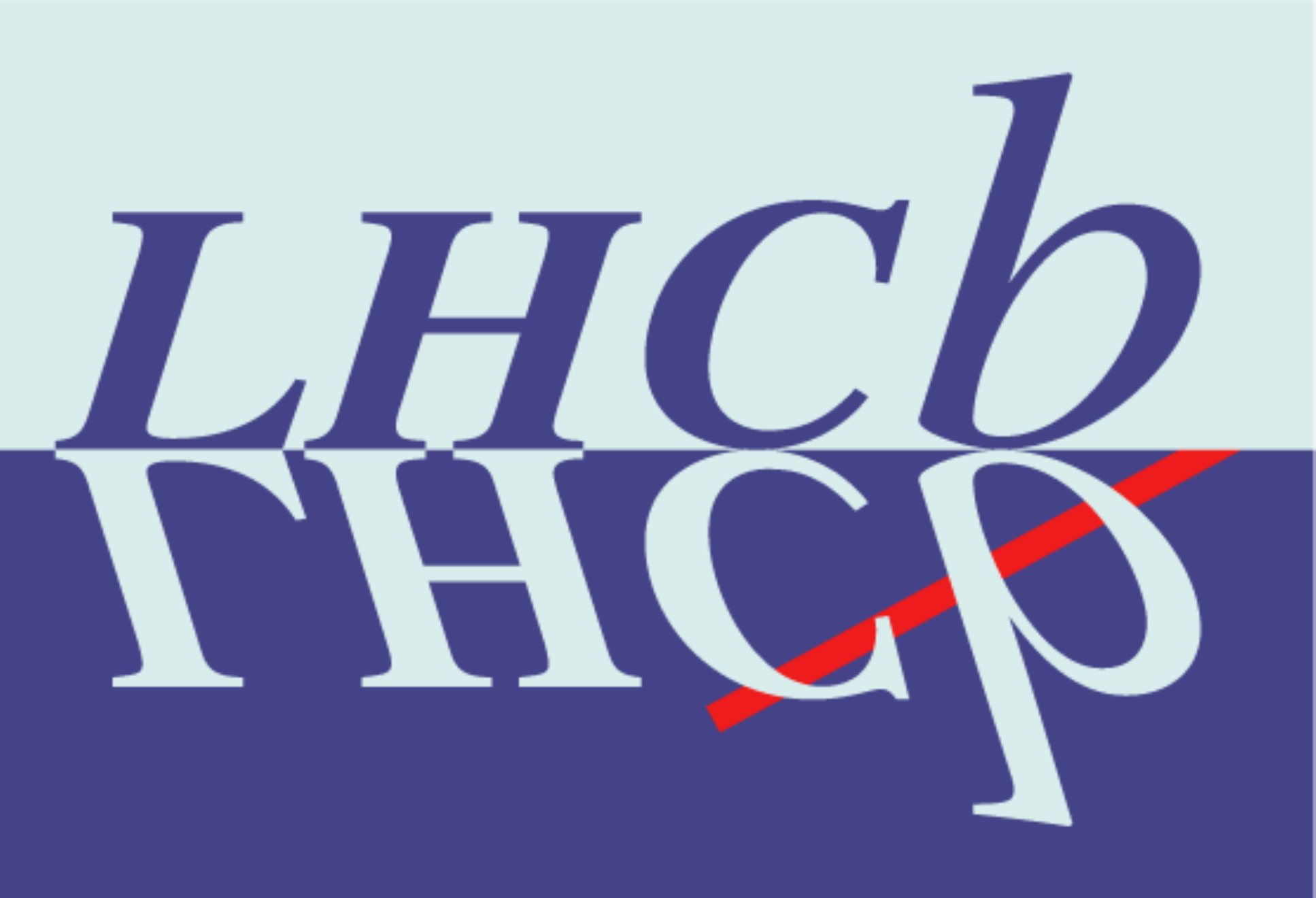}} & &}%
{\vspace*{-1.2cm}\mbox{\!\!\!\includegraphics[width=.12\textwidth]{lhcb-logo.eps}} & &}%
\\
 & & CERN-EP-2018-293 \\  
 & & LHCb-PAPER-2018-037 \\  
 & & August 20, 2019
\end{tabular*}

\vspace*{4.0cm}

{\normalfont\bfseries\boldmath\huge
\begin{center}
  \papertitle 
\end{center}
}

\vspace*{2.0cm}

\begin{center}
\paperauthors\footnote{Authors are listed at the end of this paper.}
\end{center}

\vspace{\fill}

\begin{abstract}
  \noindent
  A search for the rare leptonic decay \Bmumumu is performed using
  proton-proton collision data corresponding to an integrated
  luminosity of $4.7$ fb$^{-1}$ collected by the \lhcb experiment. The search is carried out in the
  region where the lowest of the two $\mumu$ mass combinations is
  below $980$\mevcc. The data are consistent with the background-only
  hypothesis and an upper limit of $1.6 \times 10^{-8}$ at 95\% confidence
  level is set on the branching fraction in the stated kinematic
  region.
\end{abstract}

\vspace*{2.0cm}

\begin{center}
  Published in Eur. Phys. J. C 79 
(2019) 675 
\end{center}

\vspace{\fill}

{\footnotesize
\centerline{\copyright~\papercopyright. \href{\paperlicenceurl}{\paperlicence}.}}
\vspace*{2mm}


\end{titlepage}


\newpage
\setcounter{page}{2}
\mbox{~}

\cleardoublepage


\renewcommand{\thefootnote}{\arabic{footnote}}
\setcounter{footnote}{0}



\pagestyle{plain} 
\setcounter{page}{1}
\pagenumbering{arabic}


%


\section{Introduction}
\label{sec:Introduction}

Leptonic decays of the \Bp meson are rare, as branching fractions are proportional to the squared magnitude of the small
Cabibbo-Kobayashi-Maskawa~(CKM) matrix element $V_{ub}$. Among these
processes, the decays  $B^{+}\to\taup\neut$ and $\Bp\to\mup\neum$ have
precise Standard Model~(SM) predictions~\cite{Silverman:1988gc} given
the absence of hadrons in the final state.\footnote{The inclusion of
charge-conjugate processes is implied throughout this paper.}  Due to
helicity suppression, they are also highly sensitive to particles
predicted in extensions of the SM such as charged
scalars~\cite{Isidori:2006pk}. Measurements of the $\Bp\to\taup\neut$
decay from the \B factories~\cite{Kronenbitter:2015kls, Adachi:2012mm, 
Lees:2012ju,Aubert:2009wt} lead to an average branching fraction of 
$(1.4\pm0.3)\times 10^{-4}$\cite{HFLAV16} consistent with the SM 
prediction within the experimental uncertainty. An upper limit of $1.1\times 10^{-6}$~\cite{Sibidanov:2017vph} is set on the $\Bp\to\mup\neum$ branching fraction at 90\% confidence level.

The radiative version of the muonic decay, $\Bp\to\mup\neum\gamma$, is important for two reasons; it 
is a background for the $\Bp\to\mup\neum$ decay, and its branching fraction is a direct measurement of the 
inverse moment of the \B meson light cone distribution amplitude, which
is very difficult to calculate theoretically~\cite{Beneke:2011nf}.
The upper limit on the branching fraction for the $\Bp\to\mup\neum\gamma$ decay is $3.0\times
10^{-6}$~\cite{Gelb:2018end2} at 90\%
confidence level.

\begin{figure}[ht]
  \begin{center}
    \includegraphics[width=0.45\linewidth]{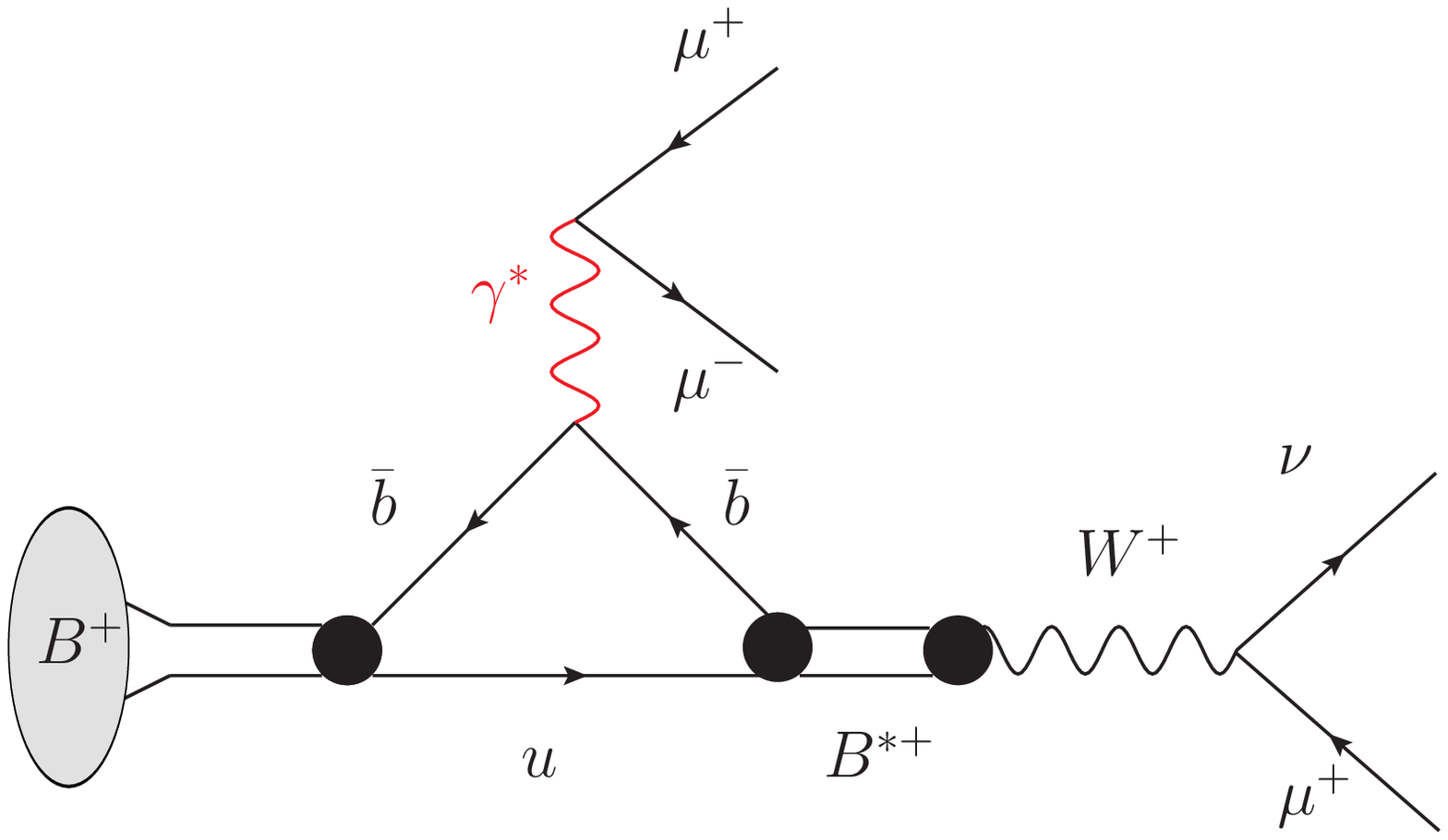}
	  \hspace*{1.0cm}
    \includegraphics[width=0.45\linewidth]{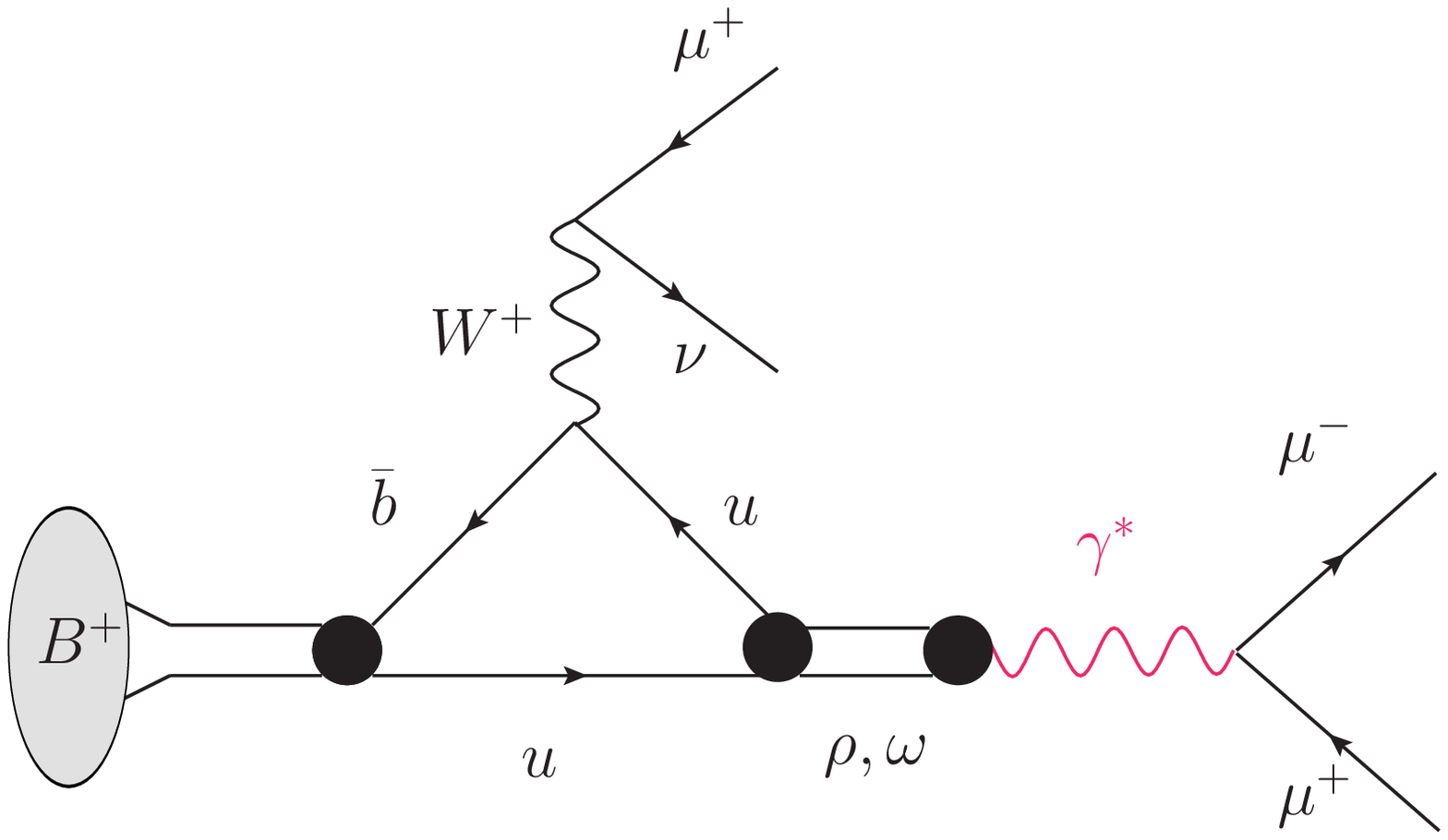}
    \vspace*{-0.5cm}
    \includegraphics[width=0.35\linewidth]{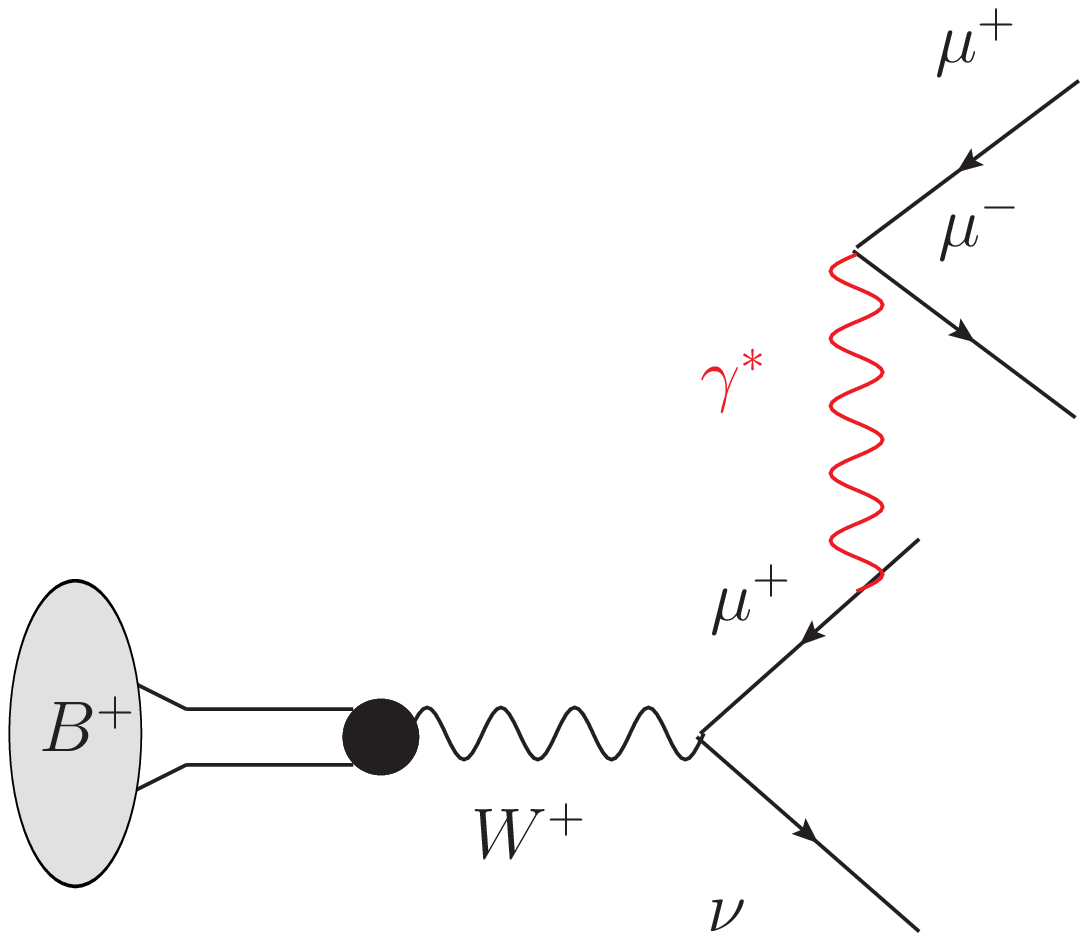}
  \end{center}
  \caption{Feynman diagrams of the contributions (top left)
    $\Bp\to\mup\neum\gamma^{*}$ with $\gamma^{*}\to\mumu$, (top right)
    $\Bp\to\mup\neum V$ and (bottom) bremsstrahlung to the \Bmumumu decay.}
  \label{fig:feyn}
\end{figure}

A $B$ decay vertex with just a single charged particle makes a search for the \mbox{$\decay{\Bp}{\mu^{+}\nu_\mu}$} and $\Bp\to\mup\neum\gamma$ decays highly challenging in the LHC environment. This problem is not present for the decay \mbox{\Bmumumu}, depicted in Fig.~\ref{fig:feyn}. The decay receives a contribution from the $\Bp\to\mup\neum\gamma^{*}$ with $\gamma^{*}\to\mumu$ amplitude, where the annihilation to the $\mup\neum$ pair occurs through an intermediate $\B^\ast$ meson. It also receives contributions from the $\Bp\to\mup\neum V$ amplitude, where $V$ denotes a vector meson such as the $\omega$ or the $\rho$, that can decay to a pair of muons. With these contributions, nearly all decays have a muon pair with a mass below 1\gevcc. The only theoretical calculation available is based on vector-meson dominance and predicts that the corresponding branching fraction, $\mathcal{B}(\Bmumumu)$, is around $1.3 \times 10^{-7}$~\cite{Danilina:2018uzr}. 

This paper describes a search for the decay \Bmumumu using a partial
reconstruction method that infers the momentum of the missing neutrino to
obtain a mass 
estimate of \Bmumumu decays. This search uses proton-proton ($pp$) collision
data corresponding to an integrated luminosity of $4.7\invfb$ collected during the  three periods 2011~(7\tev collision energy), 2012~(8\tev) and 2016~(13\tev)
at the LHCb experiment. The detector is described in Sec.~\ref{sec:Detector}, followed by a description of how
the signal is separated from backgrounds using two multivariate classifiers in Sec.~\ref{sec:Selection}. The
evaluation of the background is covered in Sec.~\ref{sec:Misid}, the normalisation of the branching fraction of
the signal to the decay $\decay{\Bp}{\jpsi \Kp}$ with $\decay{\jpsi}{\mup\mun}$ in Sec.~\ref{sec:Normalisation},
the limit on the branching fraction in Sec.~\ref{sec:results} and the systematic uncertainties in
Sec.~\ref{sec:systematics}. Finally, conclusions are presented in Sec.~\ref{sec:conclusions}.

\section{Detector and simulation}
\label{sec:Detector}

The \lhcb detector~\cite{Alves:2008zz,LHCb-DP-2014-002} is a single-arm forward spectrometer covering the \mbox{pseudorapidity} range $2<\eta <5$, designed for the study of particles containing \bquark or \cquark quarks. The detector includes a high-precision tracking system consisting of a silicon-strip vertex detector surrounding the $pp$ interaction region~\cite{LHCb-DP-2014-001}, a large-area silicon-strip detector located upstream of a dipole magnet with a bending power of about $4{\mathrm{\,Tm}}$, and three stations of silicon-strip detectors and straw drift tubes placed downstream of the magnet. The tracking system provides a measurement of the momentum, \ptot, of charged particles with a relative uncertainty that varies from 0.5\% at low momentum to 1.0\% at 200\gevc.  The minimum distance of a track to a primary vertex (PV), the impact parameter (IP), is measured with a resolution of $(15+29/\pt)\mum$, where \pt is the component of the momentum transverse to the beam, in\,\gevc.  The secondary vertex~(SV) resolution for three-body decays is around 20\mum in the plane transverse to the beam axis and 200\mum along the beam axis.

Different types of charged hadrons are distinguished using information from two ring-imaging Cherenkov detectors~\cite{LHCb-DP-2012-003}. Photons, electrons and hadrons are identified by a calorimeter system consisting of scintillating-pad and preshower detectors, an electromagnetic and a hadronic calorimeter. Muons are identified by a system composed of alternating layers of iron and multiwire proportional chambers~\cite{LHCb-DP-2012-002}.

The online event selection is performed by a multistage trigger\cite{LHCb-DP-2012-004}. For the analysis described here, the events are first required to pass a hardware trigger, selecting events containing at least one muon with high \pt. In the subsequent software trigger at least one muon candidate is required to have high \pt and a large impact parameter with respect to any PV. The dominant path through the last level of the trigger is a selection that requires a SV consisting of two muons with a high combined mass.

Simulated events are used to optimise the signal selection, estimate background contamination as well as calculate the relative efficiency between the signal and a normalisation channel. In the simulation, $pp$ collisions are generated using \pythia~\cite{Sjostrand:2006za,*Sjostrand:2007gs} with a specific \lhcb configuration~\cite{LHCb-PROC-2010-056}. Decays of hadronic particles are described by \evtgen~\cite{Lange:2001uf}, in which final-state radiation is generated using \photos~\cite{Golonka:2005pn}. The interaction of the generated particles with the detector, and its response, are implemented using the \geant toolkit~\cite{Allison:2006ve, *Agostinelli:2002hh} as described in Ref.~\cite{LHCb-PROC-2011-006}.

Three different models are used in the simulation for the \Bmumumu decay. The nominal model, with which efficiency for signal selection ($\varepsilon(\Bmumumu)$) is calculated, has a photon pole for one of the muon pairs and a uniform mass distribution for the combination of the third muon and the neutrino. For systematic checks, a flat phase space model is used. As a third model, the recently proposed vector-meson dominance model for the decay is used~\cite{Danilina:2018uzr}.

\section{Selection}
\label{sec:Selection}
Signal \Bp decay candidates are reconstructed by combining one negatively and 
two positively charged tracks. These tracks are required to be of good quality, be inconsistent with originating from any PV, be positively identified as muons and form a good-quality SV 
displaced from any PV. The PV with the smallest $\chisqip$ is the associated PV, where \chisqip\ is defined as 
the difference in the vertex-fit \chisq of a given PV reconstructed with and
without the \Bp trajectory included.
The momentum vector of the $\Bp$ decay products is 
required to point in the same direction as the line connecting the 
associated PV and the SV with an allowance made for the momentum that 
is carried away by the neutrino in the decay. 

At most one hit in the muon 
stations is allowed to be shared between two different muon candidates. This reduces the rate of hadrons misidentified as muons when there is already a muon of the same sign in the detector. In this analysis that has two muons of the same sign in
the final state it is essential to reduce this type of misidentification. The search for the signal is performed in the region where the lower of
the two $\mumu$ mass combinations is below $980$\mevcc to avoid potential 
background from 
$\decay{\phi}{\mup\mun}$ decays. Moreover, above this mass the 
combinatorial background grows and the expected signal yield is minimal, 
making a search there difficult. Backgrounds originating from candidates 
involving \jpsi and \psitwos decays are removed by vetoing the mass 
regions $2946\mevcc < M_{\mup\mun} < 3176\mevcc$ and $3586\mevcc < 
M_{\mup\mun} < 3766\mevcc$ of the higher of the two $\mumu$ mass 
combinations. Finally, a tight particle identification~(PID) selection, 
based on a neural network, is applied to reject misidentified hadrons.

The missing neutrino in the reconstruction of the \Bp candidate is accounted for with the addition of the momentum component perpendicular to \B meson flight direction, $p_\perp$. This direction is determined from the position of the PV where the \Bp meson is produced and the SV where it decays. The resulting \textit{corrected mass} is defined as,
\begin{equation}
	M_{\rm{corr}} = \sqrt{M_{\mu\mu\mu}^{2} + |p_{\perp}|^2} + |p_{\perp}|,
\end{equation}
where $M_{\mu\mu\mu}$ is the mass of the three muons. Candidates are kept if they satisfy $4000\mevcc < M_{\rm{corr}} < 7000\mevcc$. Inside this a signal region is defined as  $4500\mevcc < M_{\rm{corr}} < 5500\mevcc$. To avoid any bias in the development of the signal selection algorithm, the data in this region was not analysed until the selection was finalised and the systematic uncertainties evaluated. The uncertainty on the corrected mass is dominated by the resolution of the SV.

To reduce combinatorial background, where random tracks are combined to emulate the signal, a boosted decision tree classifier~(BDT)~\cite{Breiman} with the AdaBoost algorithm\cite{AdaBoost} as implemented in the TMVA toolkit~\cite{Hocker:2007ht,*TMVA4} is used. The BDT classifier is trained using simulation as a signal sample and the upper sideband $M_{\rm{corr}} > 5500\mevcc$ of data as a proxy of the combinatorial background candidates. To best exploit the limited amount of data available for training, a ten-fold cross-validation method~\cite{kfold} is employed. The BDT contains information about kinematic and geometric properties of the $\Bp$ candidate and associated muon tracks together with the total number of reconstructed tracks in the event. The most distinguishing properties between signal and combinatorial background candidates are the isolation of the decay vertex (as described in Ref.~\cite{LHCb-PAPER-2015-025}), the $\chi^2$ of the $\Bp$ vertex, and the \chisqip with respect to the associated PV for all three muon candidates. The requirement on the BDT response is optimised by maximising the figure of merit $\frac{\varepsilon_{S}}{\sqrt{n_B}+3/2}$~\cite{Punzi:2003bu} where $\varepsilon_{S}$ is the signal efficiency of the selection and $n_B$ refers to the estimated number of background candidates in the signal region. The optimal BDT working point is $40\%$ efficient on simulated signal events while rejecting $99\%$ of the combinatorial background. For the optimisation, only relative changes in signal efficiency are relevant and these are obtained from the simulation.

A second BDT is trained to reject contamination from misidentified
background. This background originates mostly from cascade decays where a
\bquark hadron undergoes a semileptonic decay through the dominant \bquark
to \cquark transition and the resulting \cquark hadron also decays
semileptonically. The second BDT shares the same architecture, features 
and
working-point optimisation strategy as the BDT designed to reject
combinatorial background. It is trained on a
background sample selected in data where two tracks are positively
identified as muons and the third track is required to be in the fiducial
region covered by the muon chambers but with a veto on muon
identification. The signal sample is using the simulated sample after it has been accepted by the first BDT. The optimisation results in that $40\%$ of the signal sample is retained and $94\%$ of the misidentified background is
rejected.

The overall selection results in 1797 candidates. There are no events with multiple candidates. The total efficiency for selecting the signal is about 0.1\%.

\section{Background estimation}
\label{sec:Misid}
The main categories of background are: combinatorial; misidentified combinations, where two muons are correctly identified but the third particle is a misidentified hadron; and partially reconstructed that have an almost identical final state to the signal.

As the combinatorial background arises from random combinations of three correctly identified muons, it has no peaking features in the considered region of corrected mass. Its contribution is estimated as part of the final fit to the data.

In order to estimate the number of misidentified background candidates and their distribution in the $M_{\rm{corr}}$ variable, a
data sample is obtained with the same selection as for the signal, apart from
a reversal of the muon identification requirements for one of the candidate
tracks. This track is still required to be within the fiducial volume of the
muon detector. This selects a sample of $\mu^+\mu^\pm h X$ candidates in
data, where $h$ denotes any hadron of either negative or positive charge. The
sample is a mixture of partially reconstructed \bquark-hadron decays, where
both the \bquark-hadron and the subsequent charm hadron decays semileptonically,
and combinatorial background. Backgrounds where two hadrons are identified as
muons are only contributing to the selected events at an insignificant level.

Probabilities of misidentifying hadrons as muons are obtained from data as a function of momentum and pseudorapidity by using control samples where the hadron species are determined purely from the kinematic properties of the decay chain~\cite{LHCb-DP-2012-003}. As the misidentification probability is different for pions, kaons and protons~\cite{LHCb-DP-2013-001}, the species of the hadron must be determined.  This is done by isolating the hadrons in the $\mumu h X$ sample into separate hadron PID regions and then taking into account the cross-feed, calculated using an iterative approach, between these regions. The iterative approach splits the data sample into three PID regions, where the hadron candidate is consistent with the kaon, pion and proton hypotheses, respectively. Initially, the number of misidentified candidates of a given species is assumed to be zero, and the cross-feed between regions is calculated. From this first estimate of the number of misidentified particles in each of the PID regions, the cross-feed can then be recalculated. The process repeats until the number of total misidentified particles does not change significantly from one iteration to the next when compared to the statistical uncertainty from the sample size.

Once the cross-feed between the different hadron species has been taken into account,
the probability for a specific hadron to pass the stringent muon PID requirements
applied in the analysis is calculated. The presence of the two real muons in the
$\mumu h X$ background increases the probability to misidentify the hadron as a muon,
mainly due to hit sharing in the muon stations. To take this into account, the hadron
misidentification probability is obtained using the decay $\Bz \rightarrow \jpsi \Kstarz$, with $\decay{\jpsi}{\mup\mun}$ and $\decay{\Kstarz}{\Kp\pim}$, as a calibration sample.
It has two muons present as in the signal, and the kaon and pion can be identified
without PID requirements on the particle under consideration. 
In this way the probability of identifying the kaon or the
pion from the \Kstarz decay as a muon can be measured.
Double misidentification in the calibration sample, where the kaon and pion 
hypotheses are swapped, is reduced by requiring a loose hadron identification
on the hadron not under consideration for misidentification and subsequently 
fitted for. 
The background coming from
protons misidentified as muons is insignificant, requiring no further action.

The final distribution of the misidentified background in $M_{\rm{corr}}$ is obtained by multiplying the sample with the muon identification reversed with the relevant $h\to\mu$ misidentification probabilities.

The level of partially reconstructed backgrounds, where three muons are correctly identified but one or more particles in addition to a neutrino are not reconstructed, is determined using simulation. An example of this type of decay is \mbox{${\B \rightarrow \Dzb \mup \nu_{\mu} X}$} where $\Dzb \rightarrow \Kp \pim \mup \mun$ and the $K^{+}$, $\pi^{-}$ and $X$ particles are not reconstructed. For this particular background, the measurements of the branching fractions of $\Dzb \rightarrow K^{+} \pi^{-} \mu^{+} \mu^{-}$~\cite{LHCb-PAPER-2015-043} and $\B \rightarrow \Dzb \mu^{+} \nu_{\mu} X$~\cite{PDG2018} are used. In total, partially reconstructed backgrounds are estimated at the level of eleven candidates in the signal region of corrected mass.

Other potential backgrounds are considered. The decay
$\decay{\Bp}{\Kp\mup\mun}$ with the kaon misidentified as a muon
contributes in candidates with a corrected mass outside the signal region.
This is not the case for the $\decay{\Bp}{\pip\mup\mun}$ decay, but the
low branching fraction combined with the requirement for
misindentification of the pion results in a negligible background level.
The decay $\decay{\Bp}{\eta^{(')}\mup\nu_\mu}$, followed by the decay
$\decay{\eta^{(')}}{\mup\mun\gamma}$, is also considered and found to be at a
negligible level after the selection criteria are applied.
Finally, backgrounds that involve a charmonium state decaying to a pair 
of muons are excluded by the previously mentioned vetos on the $\jpsi$ and
$\psitwos$ masses.

\section{Normalisation method}
\label{sec:Normalisation}
The branching fraction of a \Bmumumu signal is obtained by normalising to the \bjpsimumuk decay as
\begin{equation}
  \begin{aligned}
    \label{eq:BF}
	\mathcal{B}(\Bmumumu) &=\mathcal{B}(\decay{\Bp}{\jpsi\Kp}) \times
	\mathcal{B}(\decay{\jpsi}{\mup\mun}) \\
	 &\phantom{{}={}} \times
	\frac{\varepsilon(\decay{\Bp}{\jpsi\Kp})}{\varepsilon(\Bmumumu)}
	\times \frac{N(\Bmumumu)}{N(\decay{\Bp}{\jpsi\Kp})},
\end{aligned}
\end{equation}
where $N$ is the yield of the decay, $\varepsilon$ is the overall efficiency
to reconstruct and select the decay. The braching fractions are taken from
Ref.~\cite{PDG2018}.

The \decay{\Bp}{\jpsi\Kp} candidates are selected in the same way as the signal, except that
the third particle must be consistent with the kaon hypothesis and the dimuon mass
consistent with the $\jpsi$ mass. This reduces the
impact of systematic uncertainties related to the ratio of efficiencies in
Eq.~(\ref{eq:BF}). Most of the signal and normalisation selection efficiencies are
estimated using simulation. Efficiencies of the PID are obtained using control data
samples where identities of the final-state particles can be deduced from the kinematics
of the decay. The total efficiency of the \Bmumumu signal is around $37\%$ relative
to the normalisation channel. This lower efficiency is caused by the lower dimuon
mass for the signal that affects the trigger, reconstruction and BDT efficiencies.
The muon PID requirements are also less efficient due to the sharing of muon hits
between the different final-state muons in the signal decay.

\begin{figure}[t]
\centering
\includegraphics[width=0.85\linewidth]{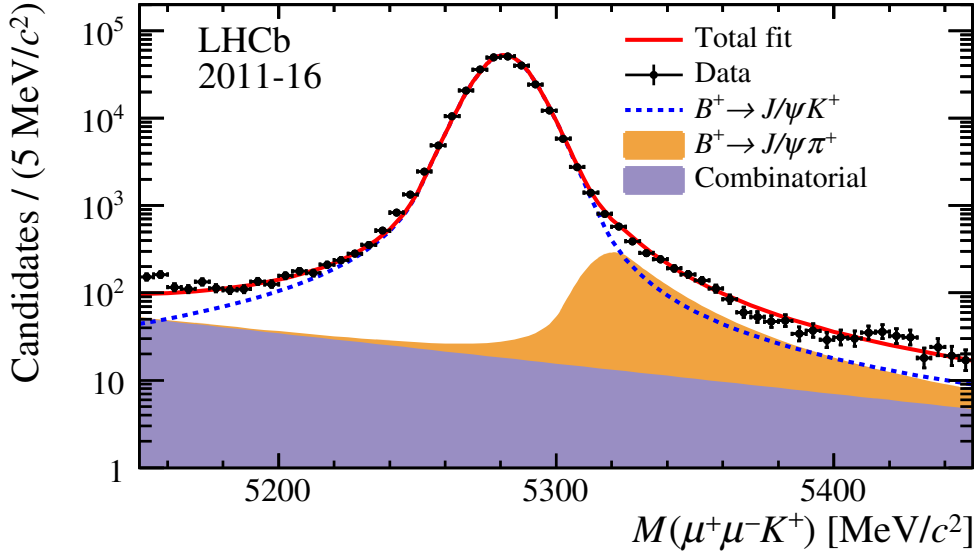}
\caption{\small Fit to the mass distribution of the selected $\decay{\Bp}{\jpsi\Kp}$ candidates. The combinatorial background (purple) and misidentified \decay{\Bp}{\jpsi\pip} decays (orange) are stacked up while the $\decay{\Bp}{\jpsi\Kp}$ signal is shown as a dashed line. The data points are shown as black points with the total fit overlaid as a red solid line.}
\label{fig:cfit}
\end{figure}
The \decay{\Bp}{\jpsi\Kp} yield is determined by performing an unbinned extended
maximum-likelihood fit to the $\mu^{+} \mu^{-} K^{+}$ mass distribution.
The shape of the \mbox{\decay{\Bp}{\jpsi\Kp}} mass distribution is described by a
Hypatia function~\cite{Santos:2013gra} that accounts for non-Gaussian tails on both
sides of the peak. In the fit, the mean and width parameters are allowed
to vary and all other parameters are determined from simulation. The 
shape of the misidentified background contribution of \decay{\Bp}{\jpsi\pip} 
decays is modelled with a Gaussian core with power law tails on each 
side of the peak. The mean and width are allowed to vary freely in 
the fit while the tail parameters are determined from simulation.
Combinatorial background is parameterised with an exponential function
with a decay constant that is allowed to vary in the fit. The result of
the fit is shown in Fig.~\ref{fig:cfit} and yields $2.7\times 10^5$
\decay{\Bp}{\jpsi\Kp} decays.

\section{Signal yield determination}
\label{sec:results}
\begin{figure}[t]
\centering
	\includegraphics[width=0.85\linewidth]{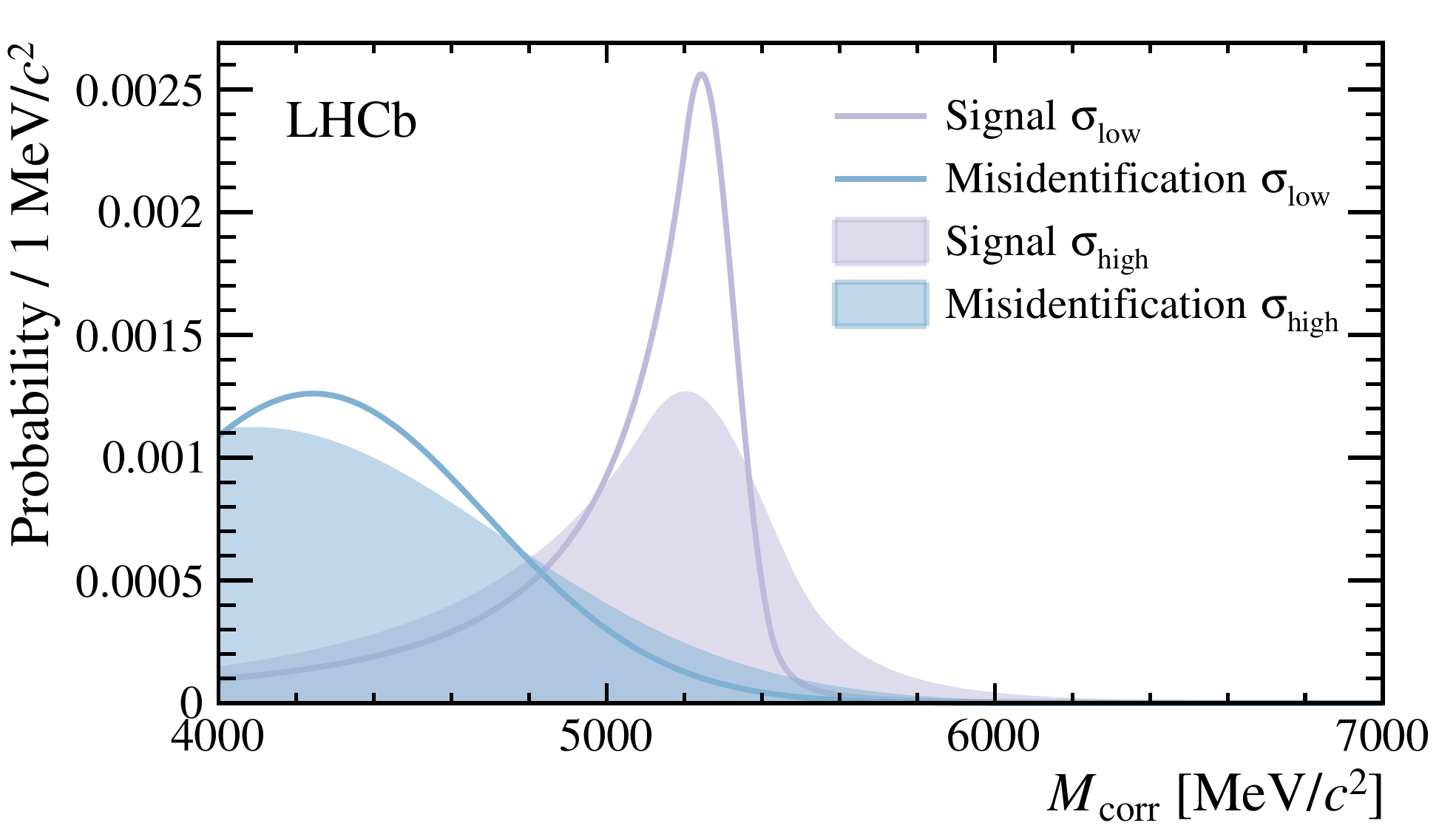}
	\caption{\small Template distributions for signal and misidentified background
          shapes for high and low fractional corrected mass
          uncertainty. A low uncertainty on the
          corrected mass corresponds to data with better mass
          resolution. The shape of the misidentification template is
          obtained from a control sample while the signal template is
          obtained from simulation. A systematic uncertainty on the signal shape due to the choice of the signal model is not shown, as it is too small to be visible.}
\label{fig:resofit}
\end{figure}

In order to determine the \Bmumumu signal yield, an extended unbinned maximum-likelihood
fit is performed to the corrected mass distribution. To improve the sensitivity of the mass fit, an event-by-event uncertainty on the corrected mass is calculated by propagating the uncertainties of the PV and SV. The data is then split into two equally sized regions with high and low fractional corrected mass uncertainties. This improves the branching fraction sensitivity by approximately 11\% due to the different signal distributions in the two samples, as shown in Fig.~\ref{fig:resofit}.

The signal shape is modelled with the sum of two Gaussian functions with power law tails, where the tails are on both sides of the peak. The parameters of the signal shape are determined using simulation and kept fixed in the subsequent fit to the data. The combinatorial background is modelled using an exponential function, whose slope is allowed to vary in the fit and whose parameterisation is verified using simulation. The yield is left free to float in the fit.

The background from misidentified muons is obtained from the $\mumu h X$ control sample described in Sec.~\ref{sec:Misid}. The distribution and yield of this sample is fitted to a Gaussian function with a power-law tail at high corrected mass. This parameterisation is cross-checked by fitting a sample with a looser muon identification requirement. The uncertainties on the associated parameters are propagated to the fit using a multivariate Gaussian constraint. The shape and the yield of the partially reconstructed background are taken from simulation. Yields that are obtained from control samples and simulation are allowed to vary in the fit within constraints from a Poisson distribution.

\begin{figure}[t]
  \centering
  \includegraphics[width=0.85\linewidth]{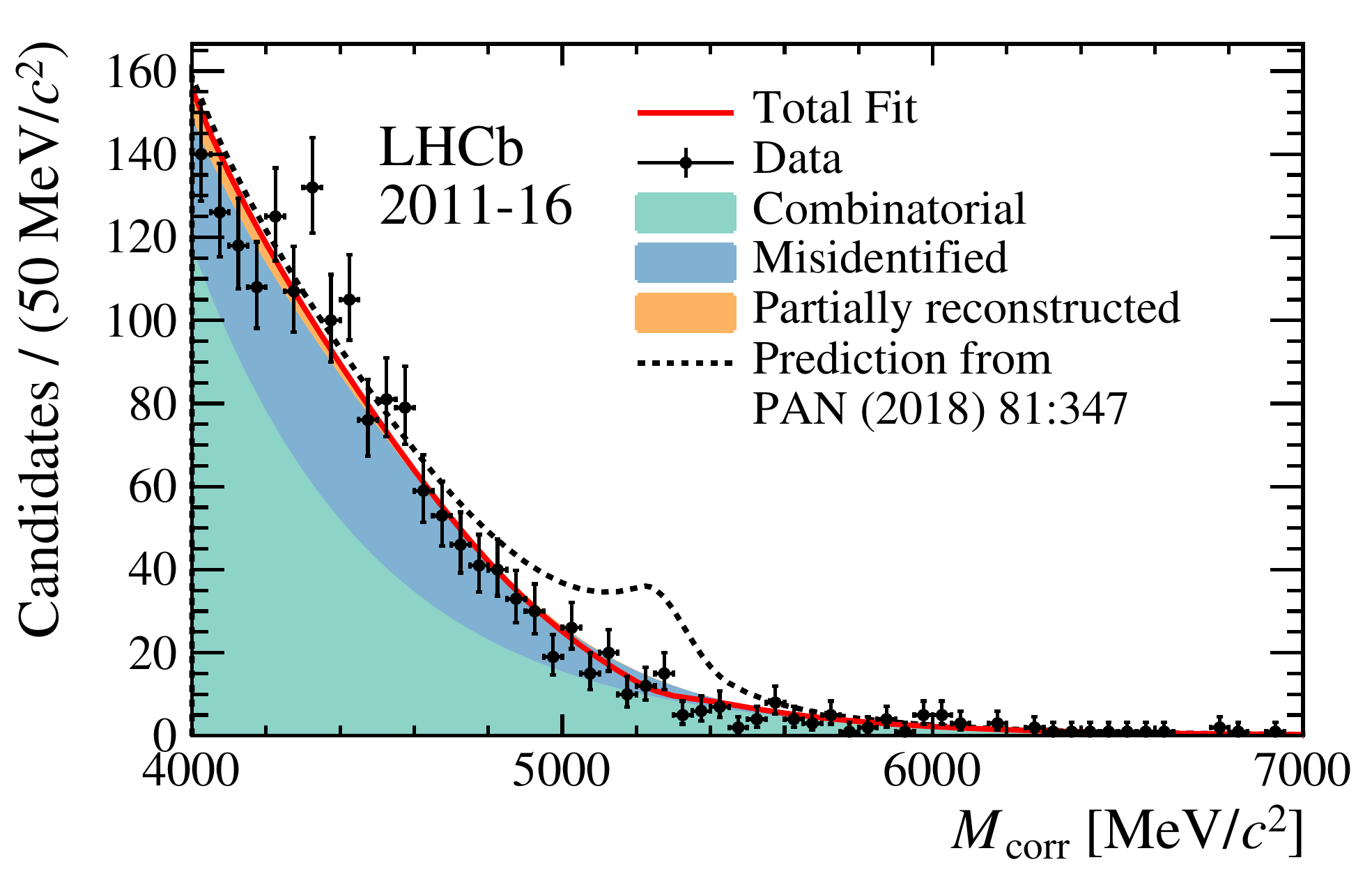}
  \caption{\small Corrected mass distribution of the selected \Bmumumu
    candidates with the fit overlaid. Samples with low and high
    corrected mass uncertainty are fitted as individual samples but
    are merged in the figure. The fit has components for (green) combinatorial background, (blue) misidentified candidates and (orange) partially reconstructed candidates. The signal component is not visible as
    the fitted signal yield is negative. The best fit is the solid red line while the dashed line shows how the
    total would have looked like if the signal had the branching
    fraction predicted in Ref.~\cite{Danilina:2018uzr}.}
  \label{fig:signalfit}
\end{figure}
The fit to the corrected-mass distribution, combining both corrected-mass uncertainty categories, is shown in Fig.~\ref{fig:signalfit}. The signal yield is negative,  $-25\pm16$, resulting in the total fit component being slightly below the sum of the background contributions. As there is no significant signal component, a limit on the branching fraction, 
\begin{equation*}
    \BF(\decay{\Bp}{\mup\mun\mup\nu_\mu}) < 1.6\times 10^{-8}
\end{equation*}
at 95\% confidence level is set using the  $\rm{CL}_{\rm s}$ method~\cite{CLS}. 
From pseudoexperiments,
the expected upper limit is found to be $2.8\times 10^{-8}$ and the present result represents a downward
fluctuation of $1.4\sigma$. Systematic uncertainties are included in this limit and are discussed in the
following section.

\section{Systematic uncertainties}
\label{sec:systematics}

A summary of the systematic uncertainties is given in Table~\ref{tab:systematicsummary}, yielding a total relative uncertainty of $16\%$ on the normalisation of the branching fraction of the signal.

The largest systematic uncertainty arises due to the choice of the shape for the combinatorial background. If the combinatorial background is allowed to have two components with different exponential slope, the upper limit on the branching fraction changes by 14.2\%. While the fit does not improve from adding in an extra component, its existence cannot be excluded from the fit to the data.

In simulation, the nominal signal model, as described in Sec.~\ref{sec:Detector}, creates a photon pole, increasing the branching fraction in the low dimuon mass region. The associated systematic uncertainty is estimated by replacing this decay with a model assuming a uniform phase-space distribution, but still with one of the muon pairs having a mass below 980\mevcc. This results in a $4.6\%$ systematic uncertainty. Using the model from Ref.~\cite{Danilina:2018uzr} results in a smaller variation.

\begin{table}[tb]
\centering
\caption{Summary of systematic uncertainties. Numbers are on the relative uncertainty of the normalisation for the branching fraction of the signal.}
\label{tab:systematicsummary}
\begin{tabular}{ l  c }
\hline
Source & Relative normalisation uncertainty [\%] \\
\hline
Combinatorial background shape & 14.2 \\
Choice of signal decay model  & \phantom{0}4.6\\
Trigger efficiency data/simulation &  \phantom{0}3.5\\
Normalisation mode branching fraction & \phantom{0}3.0 \\
Kaon interaction probability &  \phantom{0}2.0\\
Production kinematics & \phantom{0}1.5\\
Fit bias & \phantom{0}1.0  \\
Simulation sample size & \phantom{0}0.8 \\
 \hline
Total & \textbf{15.9} \\
\end{tabular}
\end{table}

Differences in simulation and data for the ratio of trigger efficiencies between the signal and normalisation channels gives rise to a systematic uncertainty as well. The effect is evaluated by comparing the difference between the trigger efficiency of \mbox{\decay{\Bp}{\jpsi\Kp}} decays in simulation and data, yielding a $3.5\%$ systematic uncertainty. This value represents a conservative estimate since it does not take into account an expected cancellation between signal and normalisation modes. The uncertainty in the branching fraction of the normalisation mode leads to a 3\% uncertainty.

Another difference between the signal and the normalisation channels is that
the kaon in the decay \decay{\Bp}{\jpsi\Kp} can undergo nuclear interactions in the detector
with a
probability proportional to the amount of material traversed and thus have a lower tracking efficiency. Following the
procedure outlined in Ref.~\cite{LHCb-DP-2013-002}, the uncertainty on this
amount of material leads to a 2\% systematic uncertainty.

Inaccuracies in the modelling of the $\Bp$ production kinematics lead to differences in efficiency between the signal and the normalisation channels. To account for this, correction weights to the \Bp meson momentum for the simulated samples are calculated using the measured distribution from \decay{\Bp}{\jpsi\Kp} decays.  The difference of 1.5\% in the relative efficiency between the signal and the normalisation channels, compared to the case where no weights are applied, is assigned as a systematic uncertainty.

Other smaller systematic uncertainties are assigned to account for a small fit bias due to the low amount of data available and the finite size of the simulation samples.

In the fit for the signal yield, all systematic uncertainties, apart from the variation in the background shape, affect the efficiency ratio and are added
as Gaussian constraints on the relevant efficiency ratios when calculating
the limit. They are assumed to be fully correlated between the bins of
fractional corrected mass uncertainty and uncorrelated between the different
effects. For the background shape, the increased freedom in the shape leads
to a larger uncertainty in the signal yield. The likelihood distribution used for determining the limit is stretched by the relative change in uncertainty around the minimum to reflect this.

\section{Conclusions}
\label{sec:conclusions}

A search has been performed for the rare leptonic decay \Bmumumu, using
4.7\invfb of proton-proton collision data collected by the \lhcb experiment.
No signal is observed for the \Bmumumu decay and an upper limit of 
$1.6\times 10^{-8}$ at 95\%~confidence level is set on the branching
fraction, where the lowest of the two $\mumu$ mass combinations is below
$980$\mevcc.
The limit for the full kinematic region stays the same under the assumption that the decay is dominated by intermediate vector mesons.

\section*{Acknowledgements}
%
%
\noindent We express our gratitude to our colleagues in the CERN
accelerator departments for the excellent performance of the LHC. We
thank the technical and administrative staff at the LHCb
institutes.
We acknowledge support from CERN and from the national agencies:
CAPES, CNPq, FAPERJ and FINEP (Brazil); 
MOST and NSFC (China); 
CNRS/IN2P3 (France); 
BMBF, DFG and MPG (Germany); 
INFN (Italy); 
NWO (Netherlands); 
MNiSW and NCN (Poland); 
MEN/IFA (Romania); 
MSHE (Russia); 
MinECo (Spain); 
SNSF and SER (Switzerland); 
NASU (Ukraine); 
STFC (United Kingdom); 
NSF (USA).
We acknowledge the computing resources that are provided by CERN, IN2P3
(France), KIT and DESY (Germany), INFN (Italy), SURF (Netherlands),
PIC (Spain), GridPP (United Kingdom), RRCKI and Yandex
LLC (Russia), CSCS (Switzerland), IFIN-HH (Romania), CBPF (Brazil),
PL-GRID (Poland) and OSC (USA).
We are indebted to the communities behind the multiple open-source
software packages on which we depend.
Individual groups or members have received support from
AvH Foundation (Germany);
EPLANET, Marie Sk\l{}odowska-Curie Actions and ERC (European Union);
ANR, Labex P2IO and OCEVU, and R\'{e}gion Auvergne-Rh\^{o}ne-Alpes (France);
Key Research Program of Frontier Sciences of CAS, CAS PIFI, and the Thousand Talents Program (China);
RFBR, RSF and Yandex LLC (Russia);
GVA, XuntaGal and GENCAT (Spain);
the Royal Society
and the Leverhulme Trust (United Kingdom);
Laboratory Directed Research and Development program of LANL (USA).



\addcontentsline{toc}{section}{References}
\setboolean{inbibliography}{true}
\bibliographystyle{LHCb}
\bibliography{main,LHCb-PAPER,LHCb-CONF,LHCb-DP,LHCb-TDR}

\newpage

\newpage
\centerline
{\large\bf LHCb collaboration}
\begin
{flushleft}
\small
R.~Aaij$^{29}$,
C.~Abell{\'a}n~Beteta$^{46}$,
B.~Adeva$^{43}$,
M.~Adinolfi$^{50}$,
C.A.~Aidala$^{77}$,
Z.~Ajaltouni$^{7}$,
S.~Akar$^{61}$,
P.~Albicocco$^{20}$,
J.~Albrecht$^{12}$,
F.~Alessio$^{44}$
% ...

{\footnotesize \it

% ...
$ ^{7}$Universit{\'e} Clermont Auvergne, CNRS/IN2P3, LPC, Clermont-Ferrand, France\\
% ...
$ ^{12}$Fakult{\"a}t Physik, Technische Universit{\"a}t Dortmund, Dortmund, Germany\\
% ...
$ ^{20}$INFN Laboratori Nazionali di Frascati, Frascati, Italy\\
% ...
$ ^{29}$Nikhef National Institute for Subatomic Physics, Amsterdam, Netherlands\\
% ...
$ ^{43}$Instituto Galego de F{\'\i}sica de Altas Enerx{\'\i}as (IGFAE), Universidade de Santiago de Compostela, Santiago de Compostela, Spain\\
$ ^{44}$European Organization for Nuclear Research (CERN), Geneva, Switzerland\\
% ...
$ ^{46}$Physik-Institut, Universit{\"a}t Z{\"u}rich, Z{\"u}rich, Switzerland\\
% ...
$ ^{50}$H.H. Wills Physics Laboratory, University of Bristol, Bristol, United Kingdom\\
% ...
$ ^{61}$University of Cincinnati, Cincinnati, OH, United States\\
% ...
$ ^{63}$Syracuse University, Syracuse, NY, United States\\
% ...
$ ^{77}$University of Michigan, Ann Arbor, United States, associated to $^{63}$\\
% ...
}
\end{flushleft}
\end{document}